\documentclass[aps,prd,onecolumn,groupedaddress,showpacs,nofootinbib,amssymb
]{revtex4}
\usepackage[dvips]{graphicx}
\usepackage{amssymb}
\usepackage{amsmath}
\usepackage{graphicx,,color}
\usepackage{amsfonts}
\usepackage{bm}
\usepackage{cancel}
\usepackage{comment}
\usepackage{color}

\allowdisplaybreaks[4]

\begin{document}

\tolerance=5000

\newcommand\be{\begin{equation}}
\newcommand\ee{\end{equation}}
\newcommand\nn{\nonumber \\}
\newcommand\e{\mathrm{e}}

\title{Topological Gravity motivated by Renormalization Group}

\author{Taisaku Mori$^1$ and Shin'ichi Nojiri$^{1,2}$\footnote{E-mail address:
nojiri@gravity.phys.nagoya-u.ac.jp}
}

\affiliation{
1. Department of Physics, Nagoya University, Nagoya
464-8602, Japan \\
\& \\
2. Kobayashi-Maskawa Institute for the Origin of Particles and
the Universe, Nagoya University, Nagoya 464-8602, Japan}

\begin{abstract}

Recently we have proposed models of topological field theory including 
gravity in 
Mod.\ Phys.\ Lett.\ A {\bf 31} (2016) no.37, 1650213 and 
Phys.\ Rev.\ D {\bf 96} (2017) no.2, 024009, 
in order to solve the problem of the cosmological constant. 
The Lagrangian densities of the models are BRS 
(Becchi-Rouet-Stora) 
exact and therefore the models 
can be regarded as topological theories. 
In the models, the coupling constants, including the cosmological constant, look as 
if they run with the scale of the universe and its behavior is very similar to 
the renormalization group. 
Motivated by these models, we propose new models with an the infrared 
fixed point, which may correspond to the late time universe, and an ultraviolet 
fixed point, which may correspond to the early universe. 
Especially we construct a model with the solutions corresponding to the 
de Sitter space-time both in the ultraviolet and the infrared fixed points. 

\end{abstract}


\maketitle

\section{Introduction}

In 
Mod.\ Phys.\ Lett.\ A {\bf 31} (2016) no.37, 1650213 \cite{Nojiri:2016mlb} 
and Phys.\ Rev.\ D {\bf 96} (2017) no.2, 024009 \cite{Mori:2017dhe}, 
models of topological field theory including gravity have 
been proposed 
in order to solve the cosmological constant problem. 
The accelerating expansion of the present universe may be generated by 
the small cosmological constant. 
Although the cosmological constant could be identified with a vacuum energy, 
the vacuum energy receives very large quantum corrections from matters and therefore
in order to obtain a realistic very small vacuum energy, very fine-tuning of the counter term 
for the vacuum energy is necessary\footnote{The discussion about the small but 
non-vanishing vacuum energy is given in \cite{Burgess:2013ara}, for example. }
Motivated {  by} this problem of large quantum corrections
 to the vacuum 
energy, models of unimodular gravity \cite{Anderson:1971pn,
Buchmuller:1988wx,Buchmuller:1988yn, Henneaux:1989zc,Unruh:1988in,
Ng:1990xz,Finkelstein:2000pg,Alvarez:2005iy,Alvarez:2006uu,
Abbassi:2007bq,Ellis:2010uc,Jain:2012cw,
Singh:2012sx,Kluson:2014esa,Padilla:2014yea,Barcelo:2014mua,
Barcelo:2014qva,Burger:2015kie,
Alvarez:2015sba,Jain:2012gc,Jain:2011jc,Cho:2014taa,Basak:2015swx,
Gao:2014nia,Eichhorn:2015bna,
Saltas:2014cta,Nojiri:2015sfd} {  have} been proposed. 
There have been also proposed many scenarios like the sequestering mechanism 
\cite{Kaloper:2013zca,Kaloper:2014dqa,Kaloper:2015jra, 
Batra:2008cc,Shaw:2010pq,Barrow:2010xt,Carballo-Rubio:2015kaa,
Tsukamoto:2017brj}. 
{  Among} of the possible scenarios, we have proposed the models 
of the topological {  field theory including gravity} 
in \cite{Nojiri:2016mlb} and the cosmology described by {  these} models has 
been discussed 
in \cite{Mori:2017dhe}. 

The large quantum corrections from {  matter} appear not only in the 
cosmological constant but other coupling constants. 
Even if we include the quantum corrections only from {  matter}, 
the following coupling constants $\alpha$, $\beta$, $\gamma$, and $\delta$ 
include {  large} quantum corrections, 
\begin{equation}
\label{A3}
\mathcal{L}_\mathrm{qc} = \alpha R + \beta R^2 
+ \gamma R_{\mu\nu} R^{\mu\nu} 
+ \delta R_{\mu\nu\rho\sigma} R^{\mu\nu\rho\sigma} \, .
\end{equation}
{  The} coefficient $\alpha$ diverges quadratically and $\beta$, 
$\gamma$, and $\delta$ diverge logarithmically.
We should note that if we include the quantum corrections from the graviton, there appear 
infinite numbers of {  divergent} quantum corrections, which is one of 
the reasons why the general relativity is not renormalizable. 
By using the formulation for the divergence in the cosmological constant proposed in 
\cite{Nojiri:2016mlb,Mori:2017dhe}, these divergences can be tuned to be finite 
\cite{Mori:2017dhe,Nojiri:2018owz}. 
In {  this} formulations, the coupling constants, $\alpha$, $\beta$, $\gamma$, 
$\delta$, 
and other coupling constants including the cosmological constant are replaced by the 
scalar fields. 
Then the divergences coming from the quantum corrections can be absorbed 
into {  a} redefinition of the scalar fields.
The fields depend on the cosmological time, or the scale of the universe. 
In this sense, the scalar fields, which corresponds to the coupling constants, 
run {  with a} scale as in the renormalization group. 
{  Motivated} by the above observation, in this paper, we propose new 
models where there appear {  an} infrared 
fixed point, which may correspond to the late time universe, and {  an} ultraviolet 
fixed point, which may correspond to the early universe. 
Especially we construct a model {  with} solutions {  connecting} 
two asymptotic 
de Sitter space-times, which correspond to the ultraviolet and the infrared fixed points. 

In the next section, we review the models of {  topological} gravity 
{  presented} in 
\cite{Nojiri:2016mlb,Mori:2017dhe,Nojiri:2018owz}. 
In Section III, we propose new models where there appear {  an} infrared 
fixed point, which may correspond to the late time universe, and {  an} ultraviolet 
fixed point, which may correspond to the early universe. 
Especially we construct a model, where the solutions expresses the flow from 
the de Sitter space-time corresponding to the ultraviolet fixed point to the 
de Sitter space-time both in the infrared fixed point. 
The last section is devoted to the summary, where we mention on the problems which 
have not been solved in this paper and some possibilities to solve {  them}
are shown in Appendix.

\section{Review of the models of topological 
field theory including gravity} 

We start to review the model proposed in \cite{Nojiri:2016mlb}. 
The action of the model is given by 
\begin{equation}
\label{CCC7} 
S' = \int d^4 x \sqrt{-g} \left\{ \mathcal{L}_\mathrm{gravity} 
+ \mathcal{L}_\mathrm{TP} \right\} + S_\mathrm{matter} \, , \quad 
\mathcal{L}_\mathrm{TP} \equiv - \lambda + \partial_\mu \lambda \partial^\mu \varphi 
 - \partial_\mu b \partial^\mu c \, .
\end{equation}
Here $\mathcal{L}_\mathrm{gravity}$ is the Lagrangian density of gravity, which may be 
arbitrary. 
The Lagrangian density $\mathcal{L}_\mathrm{gravity}$ may include 
the cosmological constant. 
In the action (\ref{CCC7}), $S_\mathrm{matter}$ is the action of matters, 
$\lambda$ and $\varphi$ are ordinary scalar fields {  while} 
$b$ is the anti-ghost field and $c$ is the ghost field. 
The (anti-) ghost fields $b$ and $c$ are fermionic (Grassmann odd) scalar.\footnote{
The action without $c$ and $b$ has been proposed 
in \cite{Shlaer:2014gna} in order to solve the problem of time. 
The cosmological perturbation in the model motivated in the model 
(\ref{CCC7}) has been investigated in \cite{Saitou:2017zyo}. } 
{  Note that no} parameter or {  coupling} constant {  appear} 
in the action (\ref{CCC7}) except in the parts of $S_\mathrm{matter}$ and 
$\mathcal{L}_\mathrm{gravity}$. 

We separate the gravity Lagrangian density $\mathcal{L}_\mathrm{gravity}$ 
into the sum of some constant $\Lambda$, which corresponds to the cosmological 
constant and may include the large quantum corrections from {  matter}, 
and the remaining part $\mathcal{L}_\mathrm{gravity}^{(0)}$ as 
$\mathcal{L}_\mathrm{gravity} = \mathcal{L}_\mathrm{gravity}^{(0)} - \Lambda$. 
By shifting the scalar field $\lambda$ by a constant $\Lambda$ 
as $\lambda \to \lambda - \Lambda$, the action (\ref{CCC7}) can be rewritten as 
\begin{equation}
\label{CCC7R}
S' = \int d^4 x \sqrt{-g} \left\{ \mathcal{L}_\mathrm{gravity}^{(0)}
 - \lambda + \partial_\mu \lambda \partial^\mu \varphi 
 - \partial_\mu b \partial^\mu c \right\} 
 -\Lambda\int d^4 x \sqrt{-g}\nabla_{\mu}\partial^{\mu}\varphi
+ S_\mathrm{matter} \, .
\end{equation}
{  Since} the cosmological constant $\Lambda$ appears as a {  coefficient of} total derivative in the action 
(\ref{CCC7R}), there is no contribution from the constant $\Lambda$ to any dynamics 
in the model. 
Thus we have succeeded to tune the large quantum corrections from {  matter} to vanish. 

As a quantum field theory, the action (\ref{CCC7}) generates {  negative} norm states 
\cite{Nojiri:2016mlb}, 
The negative norm states can be, however, removed by defining 
the physical states which are annihilated by 
the BRS (Becchi-Rouet-Stora) charge \cite{Becchi:1975nq}. 
Note that the action (\ref{CCC7}) is invariant under the {  following} infinite 
{  number} of BRS transformations, 
\begin{equation}
\label{CCC8BR}
\delta \lambda = \delta c = 0\, , \quad 
\delta \varphi = \epsilon c \, , \quad 
\delta b = \epsilon \left( \lambda - \lambda_0 \right)\, .
\end{equation}
Here $\epsilon$ is a Grassmann odd {  fermionic} parameter and 
$\lambda_0$ should {  satisfy}, 
\begin{equation}
\label{lambda0}
0 = \nabla^\mu \partial_\mu \lambda_0\, ,
\end{equation}
which is {  just} equation for $\lambda$: 
$\left( 0 = \nabla^\mu \partial_\mu \lambda\right)$ obtained by the variation 
of the action (\ref{CCC7}) with respect to $\varphi$.\footnote{
The existence of the BRS transformation where $\lambda_0$ satisfies
Eq.~(\ref{lambda0}) was pointed out by R.~Saitou.} 
In the BRS formalism, the physical states are BRS invariant and 
the unphysical states including the negative norm states are removed by 
the quartet mechanism 
proposed by Kugo and Ojima in the context of the gauge theory 
\cite{Kugo:1977zq,Kugo:1979gm}.\footnote{
We can assign the ghost number, which is conserved, $1$ for $c$ and $-1$ 
for $b$ and $\epsilon$. 
The four scalar fields $\lambda$, $\varphi$, $b$, and $c$ are called 
a quartet \cite{Kugo:1977zq,Kugo:1979gm}}
Because $\lambda - \lambda_0$ in (\ref{CCC8BR}) is 
given by the BRS transformation of the anti-ghost $b$, however, 
the BRS invariance breaks down spontaneously when $\lambda - \lambda_0$ 
does not vanish and therefore it becomes difficult 
to remove the unphysical states and keep the unitarity of the model. 
In the real universe, we find $\lambda - \lambda_0\neq 0$ in general 
because $\lambda$ plays the role of 
the dynamical cosmological constant and therefore {  BRS symmetry is spontaneously broken} in general. 
We should note, however, that in the real universe, one and only one 
$\lambda$ satisfying the equation $0 = \nabla^\mu \partial_\mu \lambda$ is realized. 
Then if we choose $\lambda_0$ to be equal to the $\lambda$ in the real universe, 
one and only one BRS symmetry in the infinite {  number} of the BRS symmetries given 
in (\ref{CCC8BR}) remains \cite{Mori:2017dhe}. 
The remaining BRS {  symmery} is enough to eliminate the unphysical states. 
and the unitarity is guaranteed. 

We can regard the Lagrangian density $\mathcal{L}_\mathrm{TP}$ in the action (\ref{CCC7}) 
as the Lagrangian density of a topological field theory proposed by Witten \cite{Witten:1988ze}. 
In {  a} topological field theory, the Lagrangian density is given by 
the BRS transformation of some quantity. 
We may consider the model where only {  one} scalar field $\varphi$ is included
but the Lagrangian density of the model vanishes identically and therefore 
the action is trivially invariant under any transformation of $\varphi$. 
Then the transformation of $\varphi$ can be regarded {  as} a gauge symmetry. 
We now fix the gauge symmetry by imposing the following gauge condition, 
\begin{equation}
\label{CCC9}
1 + \nabla_\mu \partial^\mu \varphi = 0\, .
\end{equation}
By following the procedure proposed by Kugo and Uehara \cite{Kugo:1981hm}, 
we can construct the gauge-fixed Lagrangian with the Fadeev-Popov (FP) ghost $c$ 
and anti-ghost $b$ by the BRS transformation (\ref{CCC8BR}) of 
$- b \left( 1 + \nabla_\mu \partial^\mu \varphi \right)$ by choosing $\lambda_0=0$, 
\begin{equation}
\label{SCCP2R}
\delta \left(- b \left( 1 + \nabla_\mu 
\partial^\mu \varphi \right) \right)
= \epsilon \left( - \left(\lambda - \lambda_0 \right) 
\left( 1 + \nabla_\mu \partial^\mu \varphi \right) 
+ b \nabla_\mu \partial^\mu c \right) 
= \epsilon \left( \mathcal{L} + \lambda_0 
+ \left(\mbox{total derivative terms}\right) 
\right)\, .
\end{equation}
Then we confirm that the Lagrangian density $\mathcal{L}_\mathrm{TP}$ in (\ref{CCC7}) 
is given by the BRS transformation of $- b \left( 1 + \nabla_\mu \partial^\mu \varphi \right)$ 
and the model is surely topological. 
Because $\lambda$ does not vanish in the real universe, the BRS 
invariance is broken. 
In this sense, the model (\ref{CCC7}) is not topological in the real universe, 
which could be the {  reason} why this model gives physical contributions.

The above mechanism can be applied for the divergences in (\ref{A3}) or more 
general divergences as shown in \cite{Mori:2017dhe}. 
When we consider the model in (\ref{A3}), the model in (\ref{CCC7}) is generalized as follows, 
\begin{align}
\label{A4}
\mathcal{L} =& - \Lambda - \lambda_{(\Lambda)} 
+ \left( \alpha + \lambda_{(\alpha)} \right)R 
+ \left( \beta + \lambda_{(\beta)} \right) R^2 
+ \left( \gamma + \lambda_{(\gamma)} \right) R_{\mu\nu} R^{\mu\nu} 
+ \left( \delta + \lambda_{(\delta)} \right) 
R_{\mu\nu\rho\sigma} R^{\mu\nu\rho\sigma} \nn
& + \partial_\mu \lambda_{(\Lambda)} \partial^\mu \varphi_{(\Lambda)} 
 - \partial_\mu b_{(\Lambda)} \partial^\mu c_{(\Lambda)} 
+ \partial_\mu \lambda_{(\alpha)} \partial^\mu \varphi_{(\alpha)} 
 - \partial_\mu b_{(\alpha)} \partial^\mu c_{(\alpha)} \nn
& + \partial_\mu \lambda_{(\beta)} \partial^\mu \varphi_{(\beta)} 
 - \partial_\mu b_{(\beta)} \partial^\mu c_{(\beta)} 
+\partial_\mu \lambda_{(\gamma)} \partial^\mu \varphi_{(\gamma)} 
 - \partial_\mu b_{(\gamma)} \partial^\mu c_{(\gamma)} 
+ \partial_\mu \lambda_{(\delta)} \partial^\mu \varphi_{(\delta)} 
 - \partial_\mu b_{(\delta)} \partial^\mu c_{(\delta)} \, .
\end{align}
We now shift the fields $\lambda_{(\Lambda)}$, $\lambda_{(\alpha)}$, $\lambda_{(\beta)}$, 
$\lambda_{(\gamma)}$, and $\lambda_{(\delta)}$ as follows, 
\begin{equation}
\label{A5}
\lambda_{(\Lambda)} \to \lambda_{(\lambda)} - \Lambda\, , \quad 
\lambda_{(\alpha)} \to \lambda_{(\alpha)} - \alpha\, , \quad 
\lambda_{(\beta)} \to \lambda_{(\beta)} - \beta\, , \quad 
\lambda_{(\gamma)} \to \lambda_{(\gamma)} - \gamma\, , \quad 
\lambda_{(\delta)} \to \lambda_{(\delta)} - \delta \, ,
\end{equation}
then the Lagrangian density (\ref{A4}) has the following form, 
\begin{align}
\label{A6}
\mathcal{L} =& - \lambda_{(\Lambda)} 
+ \lambda_{(\alpha)} R + \lambda_{(\beta)} R^2 
+ \lambda_{(\gamma)} R_{\mu\nu} R^{\mu\nu} 
+ \lambda_{(\delta)} R_{\mu\nu\rho\sigma} R^{\mu\nu\rho\sigma} \nn
& + \partial_\mu \lambda_{(\Lambda)} 
\partial^\mu \varphi_{(\Lambda)} 
 - \partial_\mu b_{(\Lambda)} \partial^\mu c_{(\Lambda)} 
+ \partial_\mu \lambda_{(\alpha)} \partial^\mu \varphi_{(\alpha)} 
 - \partial_\mu b_{(\alpha)} \partial^\mu c_{(\alpha)} \nn
& + \partial_\mu \lambda_{(\beta)} \partial^\mu \varphi_{(\beta)} 
 - \partial_\mu b_{(\beta)} \partial^\mu c_{(\beta)} 
+ \partial_\mu \lambda_{(\gamma)} \partial^\mu \varphi_{(\gamma)} 
 - \partial_\mu b_{(\gamma)} \partial^\mu c_{(\gamma)} 
+ \partial_\mu \lambda_{(\delta)} \partial^\mu \varphi_{(\delta)} 
 - \partial_\mu b_{(\delta)} \partial^\mu c_{(\delta)} \nn
+ \left(\mbox{total derivative terms}\right) \, . 
\end{align}
Except the total derivative terms, the obtained Lagrangian density (\ref{A6}) 
does not include the constants $\Lambda$, $\alpha$, $\beta$, $\gamma$, 
and $\delta$, which include the divergences from the quantum corrections. 
Therefore we can absorb the divergences into the redefinition of the scalar fields 
$\lambda_{(i)}$, $\left(i=\Lambda,\alpha,\beta,\gamma,\delta\right)$ and the divergences 
becomes irrelevant for the dynamics. 


In the initial model (\ref{A3}), the parameters are coupling constants but in the new models, 
(\ref{A4}) or (\ref{A6}), the parameters are replaced by dynamical scalar fields. 
This is one of the reasons why the divergence coming from the quantum corrections 
can be absorbed into the redefinition of the scalar fields. 
Furthermore because the scalar fields are dynamical, as we will see later, the scalar fields 
play the role of the running coupling constant. 


The Lagrangian density (\ref{A6}) is also invariant under 
the following BRS transformations
\begin{equation}
\label{A7}
\delta \lambda_{(i)} = \delta c_{(i)} = 0\, , \quad 
\delta \varphi_{(i)} = \epsilon c \, , \quad 
\delta b_{(i)} = \epsilon \left( \lambda_{(i)} - \lambda_{(i)0} \right)
\, , 
\quad \left(i=\Lambda,\alpha,\beta,\gamma,\delta\right)\, ,
\end{equation}
where $\lambda_{(i)0}$'s satisfy the equation, 
\begin{equation}
\label{lambda0general}
0 = \nabla^\mu \partial_\mu \lambda_{(i)0}\, ,
\end{equation}
as in (\ref{lambda0}). 
The Lagrangian density (\ref{A6}) is also given by the BRS transformation 
(\ref{A7}) with $\lambda_{(i)0}=0$, 
\begin{equation}
\label{A8}
\delta \left( \sum_{i=\Lambda,\alpha,\beta,\gamma,\delta} 
\left(- b_{(i)} \left( \mathcal{O}_{(i)} + \nabla_\mu \partial^\mu \varphi_{(i)} 
\right) \right) \right)
= \epsilon \left( \mathcal{L} 
+ \left(\mbox{total derivative terms}\right) \right)\, .
\end{equation}
As mentioned, due to the quantum correction from the graviton, 
{  an infinite number of divergences} appear. 
Let $\mathcal{O}_i$ be possible gravitational operators{  ;} then 
a further generalization of the Lagrangian density (\ref{A6}) is given by
\begin{equation}
\label{A11}
\mathcal{L} = \sum_i \left( \lambda_{(i)} \mathcal{O}_{(i)} 
+ \partial_\mu \lambda_{(i)} \partial^\mu \varphi_{(i)} 
 - \partial_\mu b_{(i)} \partial^\mu c_{(i)} \right) \, .
\end{equation}
Then all the divergences are absorbed into the redefinition of $\lambda_i$. 
The Lagrangian density (\ref{A11}) is invariant under the BRS transformation 
and given by the the BRS transformation of some quantity and therefore 
the model can be regarded as a topological field theory, again. 

{ 
As well-known. higher derivative gravity can be renormalizable but there appear 
the ghosts and therefore the higher derivative gravity model is not unitary. 
Although our model may be renormalizable because the divergence does not appear, 
the problem of the unitarity remains because the Lagrangian density (\ref{A11}) 
includes the higher derivative terms. 
In the viewpoint of the string theory, for example, we may expect that if we include 
the infinite number of higher derivative terms, the unitarity could be recovered 
but this is out of scope in this paper. 

Usually the problem of the renormalizability in quantum field theory is the predictability. 
Even if we consider the quantum theory of gravity starting from the general relativity, 
if we include an infinite number of the counterterms, 
the theory becomes finite but 
due to the infinite number of the counter terms, the model loses 
the predictability.
 In the model of (\ref{A11}), there could not be the problem of the divergence but 
because $\lambda_i$'s become dynamical, we need infinite number of the initial 
conditions or somethings and therefore even in the model (\ref{A11}), 
the predictability could be lost. 
If the $\lambda_i$'s have infrared fixed points, however, 
the predictability could be recovered. 
In the original model (\ref{A11}), however, we have not obtained non-trivial fixed points, 
which is one of the motivation why we considered the model in next section, 
where we try to construct the models with the fixed points. 
}

\section{Model Motivated by Renormalization Group}

{  We} assume that the space-time is given by the FRW 
(Friedmann-Robertson-Walker)
universe with flat spacial part {  and a scale factor $a(t)$}
\begin{equation}
\label{FRW}
ds^2 = - dt^2 + a(t)^2 \sum_{i=1}^3 \left( dx^i \right)^2 \, .{  .}
\end{equation}
Eq.~(\ref{lambda0general}) tells that the scalar fields $\lambda_{(i)}$ 
{  depend on the scale factor $a(t)$ and then become} time-dependent.
Because $\lambda_{(i)}$ correspond to the {  coupling} with the 
{  operator}
$\mathcal{O}_{(i)} $, 
Then the scale {  factor} dependence of $\lambda_{(i)}$ is similar to the
 {  scale dependence of the renormalized coupling $\lambda_{(i)}$}
Motivated by this observation, we consider the models {  with an} infrared 
fixed point, which may correspond to the late time universe, and {  an} ultraviolet 
fixed point, which may correspond to the early universe. 

We now assume the following BRS transformations instead of (\ref{CCC8BR}), 
\begin{equation}
\label{A7B}
\delta \lambda_{(i)} = \delta c_{(i)} = 0\, , \quad 
\delta \varphi_{(i)} = \epsilon c \, , \quad 
\delta b_{(i)} = \epsilon \lambda_{(i)} \, , 
\end{equation}
and consider the Lagrangian density which is given by the BRS transformation 
(\ref{A7B}) of some quantity, 
\begin{equation}
\label{A8B}
\delta \left( \sum_{i=\Lambda,\alpha,\beta,\gamma,\delta} 
\left( b_{(i)} \left( \mathcal{O}_i + \nabla_\mu \partial^\mu \varphi_{(i)} 
+ f_i\left(\lambda_{(j)} \right) \varphi_{(i)} \right) 
\right) \right)
= \epsilon \left( \mathcal{L} 
+ \left(\mbox{total derivative terms}\right) \right)\, .
\end{equation}
Here $\mathcal{O}_i$ are possible gravitational operators as in (\ref{A11}). 
and 
$f_i\left(\lambda_{(j)} \right)$'s are functions of {  $\lambda_{(j)}$}. 
Then we obtain 
\begin{equation}
\label{A11B}
\mathcal{L} = \sum_i \left( \lambda_{(i)} \mathcal{O}_{(i)} 
+ \partial_\mu \lambda_{(i)} \partial^\mu \varphi_{(i)} 
+ \lambda_{(i)} f_i\left(\lambda_{(j)} \right) \varphi_{(i)} 
 - \partial_\mu b_{(i)} \partial^\mu c_{(i)} 
 - f_i\left(\lambda_{(j)}\right) b_{(i)} c_{(i)} \right) \, .
\end{equation}
The obtained model (\ref{A11B}) is different from the original model (\ref{A3}), (\ref{A4}) 
or (\ref{A6}){ .} 
{  We are using a different gauge fixing and the background solution 
is not BRS invariant.
Then, in this background, } the model (\ref{A11B}) is not topological.

By the variation with respect to $\varphi_{(i)}$, we obtain the following equations, 
\begin{equation}
\label{CCRG2}
 - \nabla_\mu \nabla^\mu \lambda_{(i)} = \lambda_{(i)} f_i\left(\lambda_{(j)} \right) \, .
\end{equation}
In the FRW space-time with flat spacial part (\ref{FRW}), 
Eq.~(\ref{CCRG2}) can be written as follows, 
\begin{equation}
\label{CCRG3}
\frac{d^2 \lambda_{(i)}}{dt^2} + 3H \frac{d \lambda_{(i)}}{dt}
= \lambda_{(i)} f_i\left(\lambda_{(j)} \right) \, .
\end{equation}
Here $H$ is the Hubble rate defined by using the scale factor in 
Eq.~(\ref{FRW}) as $H\equiv \dot a/a$. 
By defining $\tau$ by $a=\e^\tau$, we find 
\begin{equation}
\label{CCRG4}
\frac{d}{dt} = H \frac{d}{d\tau}\, , 
\frac{d^2}{dt^2} = H^2 \frac{d^2}{d\tau^2} + \dot H \frac{d}{d\tau}\, ,
\end{equation}
and therefore we obtain 
\begin{equation}
\label{CCRG5}
H^2 \left\{ \frac{d^2 \lambda_{(i)}}{d\tau^2} + \left(3 + \frac{\dot H}{H^2} \right)
\frac{d \lambda_{(i)}}{d\tau} \right\}
= \lambda_{(i)} f_i\left(\lambda_{(j)} \right) \, .
\end{equation}
Because the change of $a$ can be identified with the scale transformation, 
we may compare (\ref{CCRG3}) with the renormalization group equation, 
\begin{equation}
\label{CCRG6}
\frac{d \lambda_{(i)}}{d\tau} = g_i\left(\lambda_{(j)} \right) \, .
\end{equation}
In {  cosmology}, the Hubble rate $H$ is usually used as energy scale 
but {  an analogy with} the renormalization group in the quantum field theory, 
{  suggest the} possibility to use the scale factor $a$ as the energy. 
{  From} 
\begin{equation}
\label{CCRG6B}
\frac{d^2 \lambda_{(i)}}{d\tau^2} = \sum_k 
\frac{\partial g_i\left(\lambda_{(j)} \right)}{\partial \lambda_{(k)}}
g_k\left(\lambda_{(j)} \right) \, ,
\end{equation}
we find 
\begin{equation}
\label{CCRG7}
f_i\left(\lambda_{(j)} \right) = \frac{H^2}{\lambda_{(i)}} 
\left\{ \sum_k 
\frac{\partial g_i\left(\lambda_{(j)} \right)}{\partial \lambda_{(k)}}
g_k\left(\lambda_{(j)} \right) 
+ \left(3 + \frac{\dot H}{H^2} \right) g_i\left(\lambda_{(j)} \right) 
\right\}\, .
\end{equation}
{  The interpretation of Eq.~(\ref{CCRG3}) as a 
renormalization group equation requires $f_i\left(\lambda_{(j)}\right)$ to 
be time independent.} 
Therefore the above identification (\ref{CCRG7}) can have any meaning only if $H$ 
is a constant at least near the fixed points, that is, the 
space-time should be, at least asymptotically, the de Sitter space-time. 
Later we consider the model where two fixed points are connected by the 
renormalization group. {  The two fixed points correspond to the ultraviolet (UV) and infrared (IR) limits}. 
Between the two fixed points, $H$ cannot be a constant because $H$ takes 
different values in the two fixed points. 
As we will see later, the scale dependence of $H$ can be absorbed into the 
redefinition of $f_i\left(\lambda_{(j)} \right)$ or $g_i\left(\lambda_{(j)}\right)$. 
We may assume that the renormalization equations (\ref{CCRG6}) has a ultraviolet or infrared 
fixed point. 
If the universe asymptotically goes to the de Sitter universe in the early time or late time. 
Then if we choose $f_i\left(\lambda_{(j)} \right)$ by (\ref{CCRG7}), the early universe 
corresponds to the ultraviolet (UV) fixed point and the late time universe to 
the infrared (IR) fixed point. 
Because the shift of $\tau$ corresponds to the change of the scale and 
$\tau$ is defined by using scale factor as $a=\e^\tau$, the UV limit corresponds to 
$\tau \to - \infty$ and therefore $a\to 0$ and the IR limit to 
$\tau \to \infty$, that is, $a\to \infty$.
In the neighborhood of the UV fixed point $\lambda^{*}_{\mathrm{UV}}$, 
we now assume,
\begin{equation}
\label{CCRG8}
\frac{dg_{(i)}\left(\lambda_{(j)}\right)}{d\lambda_{(i)}}>0 \, . 
\end{equation}
Then $g_{(i)}\left(\lambda_{(j)}\right)$ can be expressed as,
\begin{equation}
\label{CCRG9}
g_{(i)}\left(\lambda_{(j)}\right)
\approx k_{(i)\mathrm{UV}}(\lambda_{(j)})\left(\lambda_{(i)}
 -\lambda_{(i)\mathrm{UV}}\right) \, ,
\end{equation}
where $k_{(i)\mathrm{UV}}(\lambda_{(j)})$ is a function of $\lambda_{(j)}$ and 
$k_{(i)\mathrm{UV}}(\lambda_{(j)\mathrm{UV}})>0$. 
By using the approximation that $k_{(i)\mathrm{UV}}(\lambda_{(j)})$ could be regarded 
as a constant when $\lambda_{(i)} \approx \lambda_{(i)\mathrm{UV}}$, that is, 
$k_{(i)\mathrm{UV}}(\lambda_{(j)}) \approx k_{(i)\mathrm{UV}}(\lambda_{(j)\mathrm{UV}})$, 
the solution of {  (\ref{CCRG6}) with (\ref{CCRG9})} is given by 
\begin{equation}
\label{CCRG10}
\lambda_{(i)} {  \approx} \lambda_{(i)\mathrm{UV}} + \lambda_{(i)\mathrm{UV}0} 
a(t)^{k_{(i)\mathrm{UV}}(\lambda_{(j)\mathrm{UV}})} \, .
\end{equation}
Here $\lambda_{(i)\mathrm{UV}0}$ is a constant of the integration. 
On the other hand, near the IR fixed point, 
we replace $k_{(i)\mathrm{UV}}\to -k_{(i)\mathrm{IR}}$ 
and $\lambda_{(i)\mathrm{UV}}\to\lambda_{(i)\mathrm{IR}}$ 
in (\ref{CCRG9}) and (\ref{CCRG10}) as follows, 
\begin{equation}
\label{CCRG11}
g_{(i)}\left(\lambda_{(j)}\right)
{  \approx} -k_{(i)\mathrm{IR}}(\lambda_{(j)})\left(\lambda_{(i)}-\lambda_{(i)\mathrm{IR}} \right)
\, .
\end{equation}
Then we find 
\begin{equation}
\label{CCRG12}
\lambda_{(i)} \approx \lambda_{(i)\rm{IR}} + \lambda_{(i)\rm{IR}0} \left(\frac{1}{a(t)}
\right)^{k_{(i)\mathrm{IR}}\left( (\lambda_{(j)\mathrm{IR}})\right)}
\, .
\end{equation}
Here $\lambda_{(i)\mathrm{IR}0}$ is a constant of the integration. 
When $a(t)\to0$ in (\ref{CCRG10}), and $a(t)\to\infty$ in (\ref{CCRG12}), 
$\lambda_{(i)}$ goes to $\lambda_{(i)\mathrm{UV}}$ and $\lambda_{(i)\rm{IR}}$, respectively.
Thus, as long as the above condition in the neighborhood of UV (IR) fixed point 
is satisfied, $\lambda_{(i)}=\lambda_{(i)\mathrm{UV}}$ ($\lambda_{(i)}=\lambda_{(i)\rm{IR}}$) 
is surely the UV (IR) fixed point.
When $g_i\left(\lambda_{(j)}\right)$ behaves as (\ref{CCRG9}) near the UV fixed point, 
Eq.~(\ref{CCRG7}) tells that $f_i\left(\lambda_{(j)} \right)$ behaves as 
\begin{equation}
\label{CCRG7B}
f_i\left(\lambda_{(j)} \right) = \frac{H^2}{\lambda_{(i)\mathrm{UV}}} 
\left( k_{(i)\mathrm{UV}}(\lambda_{(j)\mathrm{UV}}) + 3 \right) 
k_{(i)\mathrm{UV}}(\lambda_{(j)\mathrm{UV}})\left(\lambda_{(i)}
 -\lambda_{(i)\mathrm{UV}}\right)
+ \mathcal{O} \left( \left(\lambda_{(i)}
 -\lambda_{(i)\mathrm{UV}}\right)^2 \right)\, .
\end{equation}
On the other hand, when $g_i\left(\lambda_{(j)}\right)$ behaves as (\ref{CCRG11}) 
near the IR fixed point, $f_i\left(\lambda_{(j)} \right)$ behaves as 
\begin{equation}
\label{CCRG7C}
f_i\left(\lambda_{(j)} \right) = \frac{H^2}{\lambda_{(i)\mathrm{IR}}} 
\left( k_{(i)\mathrm{IR}}(\lambda_{(j)\mathrm{IR}}) - 3 \right) 
k_{(i)\mathrm{IR}}(\lambda_{(j)\mathrm{IR}})\left(\lambda_{(i)}
 -\lambda_{(i)\mathrm{IR}}\right)
+ \mathcal{O} \left( \left(\lambda_{(i)}
 -\lambda_{(i)\mathrm{IR}}\right)^2 \right)\, .
\end{equation}

When we consider the Einstein gravity with cosmological constant, the action is given by,
\begin{equation}
\label{CCRG16}
S=\int d^{4}x\sqrt{-g} \left[
\lambda_{(\alpha)} R-\lambda_{(\Lambda)}
+\sum_{i=\Lambda,\alpha}
\left(
\partial_{\mu}\lambda_{(i)}\partial^{\mu}\varphi_{(i)}
-\partial_{\mu}b_{(i)}\partial^{\mu}c_{(i)}
+\lambda_{(i)}f_{(i)}(\lambda_{(j)})\varphi_{(i)}
\right)
\right]
+S_\mathrm{matter}
\,.
\end{equation}
Here $S_\mathrm{matter}$ is the action of matters. 
Varying the action (\ref{CCRG16}) with respect to the metric $g^{\mu\nu}$, we obtain the 
following equation, 
\begin{equation}
\label{CCRG17}
\lambda_{(\alpha)}G_{\mu\nu}
+\frac{1}{2}\lambda_{(\Lambda)} g_{\mu\nu} 
 - \left( \nabla_\mu \nabla_\nu - \nabla^2 \right) \lambda_{(\alpha)}
+\sum_{i=\Lambda,\alpha}
\left[
\frac{1}{2}g_{\mu\nu}
\left(
\partial_{\rho}\lambda_{(i)}\partial^{\rho}\varphi_{(i)}
+\lambda_{(i)}f_{(i)}(\lambda_{(j)})\varphi_{(i)}
\right)
+\partial_{\mu}\lambda_{(i)}\partial_{\nu}\varphi_{(i)}
\right]
=T_{\mu\nu}
\, .
\end{equation}
We should note that if the FP ghost and anti-ghost has any classical 
value, which may correspond to the vacuum expectation value, superselection rule 
or ghost number conservation is violated and therefore we put them vanish. 
In (\ref{CCRG17}), $G_{\mu\nu}$ is the Einstein tensor and $T_{\mu\nu}$ is 
the energy momentum tensor of matters. 
In the spatially flat FRW background if we assume {  that} $\lambda_{(i)}$ and $\varphi_{(i)}$ 
depend {  only} on the cosmological time $t$, the $(0,0)$-component of Eq.~(\ref{CCRG17}) has the following form, 
\begin{equation}
\label{CCRG19}
H^{2}=\frac{1}{6\lambda_{(\alpha)}} 
\left\{
\lambda_{(\Lambda)} - 3 H \dot{\lambda}_{(\alpha)} 
 - \sum_{i=\Lambda,\alpha} \left(
\dot{\lambda}_{(i)}\dot{\varphi}_{(i)} - \lambda_{(i)}f_{i}(\lambda_{j})\varphi_{(i)}
\right)
\right\}
\end{equation}
In the the neighborhood of the UV fixed point, substituting 
(\ref{CCRG7}) and (\ref{CCRG10}) into the above expression, we obtain,
\begin{align}
\label{CCRG20}
H^{2} \approx &
\frac{1}{6\lambda_{(\alpha)}}
\left( \lambda_{(\Lambda)\mathrm{UV}} + \lambda_{(\Lambda)\mathrm{UV}0} 
a(t)^{k_{(\Lambda)\mathrm{UV}}(\lambda_{(j)\mathrm{UV}})} 
 - 3 H \dot{\lambda}_{(\alpha)}
+\sum_{i=\Lambda,\alpha}k_{(i)}(\lambda_{(j)}) 
H a(t)^{k_{(i)}(\lambda_{(j)})}\lambda_{(i)\mathrm{UV}0}\dot{\varphi}_{(i)}
\right) 
\nn
&+\frac{H^{2}}{6\lambda_{(\alpha)}}
\sum_{i=\Lambda,\alpha}
\left\{ \sum_k 
\frac{\partial g_i\left(\lambda_{(j)} \right)}{\partial \lambda_{(k)}}
g_i\left(\lambda_{(j)} \right) 
+ \left(3 + \frac{\dot H}{H^2} \right) g_i\left(\lambda_{(j)} \right) 
\right\}
\varphi_{(i)}
\,,
\end{align}
Then in the UV limit
\begin{align}
\label{CCRG21}
a(t)\to 0\, ,\quad g_{(i)}\to 0 \, , \quad 
\lambda_{(i)} \to \lambda_{(i)\mathrm{UV}}
\,,
\end{align}
we obtain the de-Sitter solution, where $H$ is a constant, 
\begin{align}
\label{CCRG22}
H=H_{\mathrm{UV}}=\sqrt{\frac{\lambda_{(\Lambda)\mathrm{UV}}}
{6\lambda_{(\alpha)\mathrm{UV}}}}=\mathrm{const.}
\end{align}
On the other hand, near the IR fixed point, instead of (\ref{CCRG20}), we obtain 
\begin{align}
\label{CCRG23}
H^{2} \approx & 
\frac{1}{6\lambda_{(\alpha)}}
\left( 
\lambda_{(\Lambda)\rm{IR}} + \lambda_{(\Lambda)\rm{IR}0} a(t)
^{-k_{(\Lambda)\mathrm{IR}}\left( \lambda_{(j)\mathrm{IR}}\right)}
 - 3 H \dot{\lambda}_{(\alpha)}
 -\sum_{i=\Lambda,\alpha}k_{(i)}\left( \lambda_{(j)} \right) 
Ha(t)^{-k_{(i)}\left( \lambda_{(j)} \right)} \lambda_{(i)\mathrm{IR}0}
\dot{\varphi}_{(i)} 
\right)
\nn
&+\frac{H^{2}}{6\lambda_{(\alpha)}}
\sum_{i=\Lambda,\alpha}
\left\{ \sum_k 
\frac{\partial g_i\left(\lambda_{(j)} \right)}{\partial \lambda_{(k)}}
g_i\left(\lambda_{(j)} \right) 
+ \left(3 + \frac{\dot H}{H^2} \right) g_i\left(\lambda_{(j)} \right) 
\right\}
\varphi_{(i)}
\,,
\end{align}
Then in the IR limit
\begin{align}
\label{CCRG24}
a(t)\to\infty\, , \quad g_{(i)}\to 0 \, , \quad 
\lambda_{(i)} \to \lambda_{(i)\mathrm{IR}}
\,,
\end{align}
we obtain the de-Sitter solution, where 
\begin{align}
\label{CCRG25}
H=H_{\rm{IR}}=\sqrt{\frac{\lambda_{(\Lambda)\mathrm{IR}}}{6\lambda_{(\alpha) \mathrm{IR}}}}
=\mathrm{const.}
\end{align}

We now try to construct a model, where the IR fixed point is connected with the 
UV fixed point by the renormalization flow. 
As an example, we may consider the following model 
\begin{equation}
\label{CCRG28B}
f_{(i)}\left(\lambda_{(j)} \right) = C_{(i)} \left(\lambda_{(j)} \right)
\left(\lambda_{(i)}-\lambda_{(i)\mathrm{UV}}\right)
\left(\lambda_{(i)}-\lambda_{(i)\rm{IR}}\right)
\, ,
\end{equation}
Here $C_{(i)} \left(\lambda_{(j)} \right)$ is a positive function. 
By using (\ref{CCRG22}) and comparing (\ref{CCRG7B}) and (\ref{CCRG28B}), 
we find 
\begin{equation}
\label{CCRGA1}
\frac{\lambda_{(\Lambda)\mathrm{UV}}}
{6\lambda_{(\alpha)\mathrm{UV}}\lambda_{(i)\mathrm{UV}}}
\left( k_{(i)\mathrm{UV}}(\lambda_{(j)\mathrm{UV}}) + 3 \right) 
k_{(i)\mathrm{UV}}(\lambda_{(j)\mathrm{UV}})
= C_{(i)} \left(\lambda_{(j)\mathrm{UV}} \right)
\left(\lambda_{(i)\mathrm{UV}}-\lambda_{(i)\rm{IR}}\right) \, ,
\end{equation}
which can be solved with respect to $k_{(i)\mathrm{UV}}>0$, as follows,
\begin{equation}
\label{CCRGA2}
k_{(i)\mathrm{UV}} = - \frac{3}{2} + \frac{1}{2} 
\sqrt{ 9 + \frac{24 \lambda_{(\alpha)\mathrm{UV}}\lambda_{(i)\mathrm{UV}}
C_{(i)} \left(\lambda_{(j)\mathrm{UV}} \right)}
{\lambda_{(\Lambda)\mathrm{UV}}}\left(\lambda_{(i)\rm{UV}}-\lambda_{(i)\rm{IR}}\right)} \, .
\end{equation}
On the other hand, by using (\ref{CCRG25}) and comparing (\ref{CCRG7C}) and (\ref{CCRG28B}), 
we find 
\begin{equation}
\label{CCRGA3}
\frac{\lambda_{(\Lambda)\mathrm{IR}}}
{6\lambda_{(\alpha)\mathrm{IR}}\lambda_{(i)\mathrm{IR}}}
\left( k_{(i)\mathrm{IR}}(\lambda_{(j)\mathrm{IR}}) - 3 \right) 
k_{(i)\mathrm{IR}}(\lambda_{(j)\mathrm{IR}})
= - C_{(i)} \left(\lambda_{(j)\mathrm{IR}} \right)
\left(\lambda_{(i)\mathrm{UV}}-\lambda_{(i)\rm{IR}}\right) \, ,
\end{equation}
which can be solved with respect to $k_{(i)\mathrm{IR}}>0$, as follows,
\begin{equation}
\label{CCRGA4}
k_{(i)\mathrm{IR}} = \frac{3}{2} \pm \frac{1}{2} 
\sqrt{ 9 - \frac{24 \lambda_{(\alpha)\mathrm{IR}}\lambda_{(i)\mathrm{IR}}
C_{(i)} \left(\lambda_{(j)\mathrm{IR}} \right)}
{\lambda_{(\Lambda)\mathrm{IR}}}\left(\lambda_{(i)\mathrm{UV}}-\lambda_{(i)\rm{IR}}\right) } \, ,
\end{equation}
which requires 
\begin{equation}
\label{CCRGA5}
9 \geq \frac{24 \lambda_{(\alpha)\mathrm{IR}}\lambda_{(i)\mathrm{IR}}
C_{(i)} \left(\lambda_{(j)\mathrm{IR}} \right)}
{\lambda_{(\Lambda)\mathrm{IR}}}\left(\lambda_{(i)\mathrm{UV}}-\lambda_{(i)\rm{IR}}\right) 
 \, .
\end{equation}
Therefore as long as we choose $C_{(i)} \left(\lambda_{(j)} \right)$ to 
satisfy the constraint (\ref{CCRGA5}), the model (\ref{CCRG28B}) surely 
connect the IR fixed point with the UV fixed point by the renormalization flow.

\section{Summary}

Motivated with the model in \cite{Nojiri:2016mlb,Mori:2017dhe,Nojiri:2018owz}, 
we have proposed models of topological {  field theory including gravity}. 
In {  those models}, the coupling constants are replaced by scalar fields, which run as in 
the renormalization group {  following} the scale of the universe. 
As an example, we have constructed a model which connects the inflation in the early 
universe and the accelerating expansion of the present universe or late time. 
The de Sitter space-times corresponding to the inflation and the late time 
accelerating expansion appear as the ultraviolet and infrared fixed points, respectively. 
There remains, however, several problems, which violate the good properties in 
the original models in \cite{Nojiri:2016mlb,Mori:2017dhe,Nojiri:2018owz}. 
\begin{enumerate}
\item Because the shift symmetry as in (\ref{A4}) is lost, the models in this paper do 
not solve the problem of the large quantum correction. 
\item Because $\lambda_{(i)}$ in (\ref{A7}) has a non-trivial value, the BRS 
symmetry in (\ref{A7}) should be broken. 
\item Although the original model in \cite{Nojiri:2016mlb,Mori:2017dhe,Nojiri:2018owz} has 
no parameters, the models proposed in this paper should {  have} several parameters. 
\end{enumerate}
Therefore it could be interesting if we construct any model which solve some of 
the above problems by keeping the structure similar to the renormalization group. 
Some {  ideas to try to solve these problems} problems are given in Appendix. 

In summary, {  we} have not succeeded to solved all the problems but we may have 
shown that there might be possibilities to solve them.
In this paper, we have considered {  models} where the scalar fields 
$\lambda_{(i)}$'s play the role of the running coupling constants as in the renormalization 
group. 
We have treated the scalar fields classically although the renormalization group, of course, 
comes from the quantum corrections. 
Therefore the models proposed in this paper might be realized by the effective field theory 
connecting the low energy region with the high energy regions. 
If the models are really given as effective theories, the models need not always 
to satisfy all the unitarity conditions. 

{ 
We have anyway succeeded to construct such models and we have shown that 
we can construct the model with fixed point. 
The models have, however, arbitrariness, which could be removed by 
the constraints from the observations and/or the consistencies of the models. 
We like to reserve the problem in the future work. 
}

\section*{Acknowledgments.}

This work is supported (in part) by 
MEXT KAKENHI Grant-in-Aid for Scientific Research on Innovative Areas ``Cosmic
Acceleration'' No. 15H05890 (S.N.) and the JSPS Grant-in-Aid for
Scientific Research (C) No. 18K03615 (S.N.). 

\appendix

\section{Some propositions to improve the models}

In this appendix, we consider models, which may solve the problem given in Summary Section. We believe the models in this section may give some clues to solve the problems. 

An example of the model, which may solve the second problem, could be 
\begin{align}
\label{A8BB}
& \delta \left( \sum_{i=\Lambda,\alpha,\beta,\gamma,\delta} 
\left( b_{(i)} \left( \mathcal{O}_i + \nabla_\mu \partial^\mu \varphi_{(i)} 
\pm k_{(0)i} \varphi_{(i)} \right) \right) \right)
= \epsilon \left( \mathcal{L} 
+ \left(\mbox{total derivative terms}\right) \right)\, , \nn
& \mathcal{L} = \sum_i \left( \lambda_{(i)} \mathcal{O}_{(i)} 
+ \partial_\mu \lambda_{(i)} \partial^\mu \varphi_{(i)} 
\pm k_{(0)i} \lambda_{(i)} \varphi_{(i)} 
 - \partial_\mu b_{(i)} \partial^\mu c_{(i)} 
\mp k_{(0)i} b_{(i)} c_{(i)} \right) \, .
\end{align}
Then $\lambda_{(i)}=0$ is a ultraviolet (infrared) fixed point for $+k_{(0)i}$ 
$\left( - k_{(0)i} \right)$. 
By the variation of $\varphi_{(i)}$, we obtain 
\begin{equation}
\label{A11BB1}
0 = - \nabla^\mu \partial_\mu \lambda_{(i)} 
\pm k_{(0)i} \lambda_{(i)} \, .
\end{equation}
Let a solution of (\ref{A11BB1}) be $\lambda_{(i)}=\lambda_{(i)}^\mathrm{cl}$. 
Then the action given by the Lagrangian density $\mathcal{L}$ in (\ref{A8BB}) is 
invariant under the following BRS transformation instead of (\ref{A7B}), 
\begin{equation}
\label{A7BB}
\delta \lambda_{(i)} = \delta c_{(i)} = 0\, , \quad 
\delta \varphi_{(i)} = \epsilon c \, , \quad 
\delta b_{(i)} = \epsilon \left( \lambda_{(i)} - \lambda_{(i)}^\mathrm{cl} \right)\, , 
\end{equation}
Then because one of the solutions in $\lambda_{(i)}^\mathrm{cl}$ is realized in the real 
world, the BRS symmetry corresponding to the solution $\lambda_{(i)}^\mathrm{cl}$ 
is not broken and the unitarity can be preserved. 

Another kind of the solution may be given by the following kind of model, 
\begin{equation}
\label{CCK1}
S = \int d^4 x \sqrt{-g} \left\{ \frac{R}{2\kappa^2} - \lambda 
+ \mathcal{L} \left( g_{\mu\nu}, X, Y_{\mu\nu} \right) \right\} \, , \quad 
X \equiv - \partial_\mu \lambda \partial^\mu \lambda \, , \quad 
Y_{\mu\nu} \equiv \nabla_\mu \partial_\nu \lambda \, .
\end{equation}
Here $\mathcal{L}$ could be the Lagrangian density of the $k$-essence or 
the Galileon model. 
Because $\mathcal{L}$ is invariant under the shift of $\lambda$ by a constant $\lambda_0$: 
$\lambda \to \lambda + \lambda_0$, the vacuum energy can be absorbed into the definition 
of $\lambda$ and the first problem could be solved. 
Then if we choose $\mathcal{L}$ to give a unitary model, we need not to consider the second 
problem. 
When we consider $\mathcal{L}$ of the $k$-essence, $\mathcal{L} = \mathcal{L}(X)$, 
for simplicity, by the variation of $\lambda$, we obtain
\begin{equation}
\label{CCK3}
0 = 1 - 2 \nabla^\mu \left( \partial_\mu \lambda \mathcal{L}' \left( X \right) \right) \, .
\end{equation}
In the FRW universe with the flat spacial part (\ref{FRW}), Eq.~(\ref{CCK3}) has 
the following form,
\begin{equation}
0 = 1 + 2 a(t)^{-3} \frac{d}{dt} \left( a(t)^3 \dot \lambda 
\mathcal{L}' \left( { \dot \lambda }^2 \right) \right) \, ,
\end{equation}
which tells that the fixed point, where $\dot\lambda=0$ is not the solution.

\end{document}